\documentclass{webofc}
\usepackage[varg]{txfonts}   
\begin{document}
\title{Proton Holography}
\subtitle{Discovering Odderon from  Scaling Properties of Elastic Scattering }
%
%

\author{
        \lastname{T. Cs\"org\H{o}}\inst{1,2}\fnsep\thanks{\email{tcsorgo@cern.ch}} \and
        \lastname{T. Nov\'ak}\inst{2}\fnsep\thanks{\email{tamas.novak@uni-eszterhazy.hu}} \and
        \lastname{R. Pasechnik}\inst{3,4}\fnsep\thanks{\email{Roman.Pasechnik@thep.lu.se}} \and
        \lastname{A. Ster}\inst{1}\fnsep\thanks{\email{Ster.Andras@wigner.mta.hu}} \and
        \lastname{I. Szanyi}\inst{1,5}\fnsep\thanks{\email{Istvan.Szanyi@cern.ch}}
}

\institute{
        Wigner RCP, H-1525 Budapest 114, P.O.Box 49, Hungary
\and
        Eszterh\'azy University KRC, H-3200 Gy\"ongy\"os, Mátrai \'ut 36, Hungary
\and
        Department of Astronomy and Theoretical Physics, Lund University, SE-223 62 Lund, Sweden
\and
        Nuclear Physics Institute ASCR, 25068 Rež, Czech Republic
\and
        E\"otv\"os University, H - 1117 Budapest XI, P\'azm\'any P. s. 1/A, Hungary
          }

\abstract{%
 We investigate the scaling properties of elastic scattering data at ISR and LHC energies,
and find that the significance of an Odderon observation is  larger than the discovery threshold 
of 5$\sigma$. As an unexpected by-product of these investigations, for certain experimentally 
relevant cases, we also conjecture the possibility of proton holography with the help of elastic 
proton-proton scattering.
}
\maketitle
\section{Introduction}
\label{s:intro}
This work summarizes two related topics:
the discovery of a significant Odderon effect at LHC energies and  the investigation of 
the possibility of four-momentum transfer $t$ dependent phase measurement in elastic 
proton-proton ($pp$) and proton-antiproton ($p\bar p$) scattering.

The Odderon corresponds to a crossing-odd contribution to the scattering amplitude of elastic $pp$ 
and $p\bar p$ scattering at asymptotically high energies, proposed by Lukaszuk and Nicolescu in
1973~\cite{Lukaszuk:1973nt}. In QCD, the quantum field theory of the strong interactions, such an 
Odderon exchange corresponds to the $t$-channel exchange of a color-neutral gluonic compound state 
consisting of an odd number of gluons, as noted by Bartels, Vacca and Lipatov in 1999~\cite{Bartels:1999yt}.

As the modulus square of the elastic scattering amplitude is proportional to the 
differential cross-section of elastic scattering, a phase reconstruction is equivalent 
to the possibility of proton holography. It is well known that such a $t$-dependent phase 
reconstruction from the measurements of modulus squared amplitudes, without further information, 
is simply impossible. We thus ask a different question here: what are those 
specific additional, and experimentally testable conditions, that actually allow for a $t$-dependent 
phase reconstruction in elastic $pp$ collisions at LHC energies?

Holography of light by now is well developed technique that has several sub-topics and applications 
not only in science but also in arts, banking, programming, interferometry and security, including
holograms on bank-notes, vehicles, credit and identity cards as well as passports. 
The essential point of holography was highlighted in D. Gabor's Nobel Lecture~\cite{Gabor:1971nbl}: 
Holography is based on the wave nature of light, and corresponds to phase level reconstruction.
As every quantum field has a dual wave and quantum property, holography is possible not only using 
the wave properties of light quanta, but also
that of other particles like electrons~\cite{doi:10.1002/jemt.20098}, atoms~\cite{PhysRevLett.88.123201} 
as well as  neutrons~\cite{Sarenac:16}.
The key concept of holography is to produce a coherent source of these quanta, and have parts 
of this field diffractively scatter
on some scattering center. Recording the resulting interference pattern corresponds to recording 
a holographic picture. Illuminating this picture
with the original beam allows for reconstruction of the scattered wave including both the modulus 
and the phase of the amplitude. Actually, two different holographic images are created, the object 
beam that carries the original phase and modulus, and the conjugated object beam that reconstructs 
the modulus but conjugates the complex phase of the object beam.

The title of this work is proton holography, as we discuss here possibilities for a phase-level 
reconstruction using the elastically scattered protons at the TeV energy scale.
Although such a title may look a bit fancy, it turns out that our idea is not unprecedented, 
as Ref.~\cite{Nagy:1978iw} has already considered four-momentum transfer dependent phase reconstruction 
in elastic scattering of protons at the ISR energies of $\sqrt{s}= 23.5 - 62.5 $ GeV, but without 
introducing such a term.

A related topic is the search for a crossing-odd contribution in elastic proton-proton collisions
at CERN LHC energies, the so called Odderon effect. As the Odderon effect is not significant 
at the ISR energy range but is found to be significant at LHC energies~\cite{Csorgo:2019ewn}, 
the methodology of phase reconstruction at these two energies are also slightly different. 
For example, the phase-level reconstruction of Ref.~\cite{Nagy:1978iw}
disregarded any Odderon contribution, so that method cannot be readily applied at TeV energies.
Due to this reason, let us first briefly summarize some of our new results related to the Odderon 
discovery at the CERN LHC, and this way also prepare the ground for the idea of proton 
holography at high energies, as detailed in the second part of this work.

\section{Formalism}
\label{s:formalism}
Let us consider the elastic scattering of particles $a$ and $b$ with incoming four-momenta $(p_1,p_2)$, 
and outgoing four-momenta $(p_3,p_4)$, respectively. The Mandelstam variables $s$ and $t$ are defined as 
$s = (p_1 + p_2)^2$ and $t = (p_1 - p_3)^2$. The differential cross-section of such $(a,b)\rightarrow (a,b)$
scattering, $\frac{d\sigma(s,t)}{dt}$ can be expressed in terms of the scattering amplitude 
$T_{\rm el}(s,t)$ as
\begin{equation}
\frac{d\sigma(s,t)}{dt}   =   \frac{1}{4\pi}|T_{\rm el}(s,t)|^2 \, .
\label{e:dsigmadt-Tel}
\end{equation}
The elastic cross-section is given as an integral of the differential cross-section as
\begin{equation}
     \sigma_{\rm el}(s) = \int_{0}^\infty {d}t \,\, \frac{d\sigma(s,t)}{dt} .
     \label{e:sigmael}
\end{equation}

At a given $s$, the $t$-dependent slope parameter $B(s,t)$ is the logarithmic slope of the differential cross-section:
\begin{equation}
    B(s,t) = \frac{d}{dt} \ln \frac{d\sigma(s,t)}{dt} \,. 
    \label{e:Bst}
\end{equation}
In the low-$t$ region, corresponding to the diffractive cone, this function is frequently assumed 
or found within experimental errors to be a constant. To characterize this diffractive cone,
a $t$-independent slope parameter $B(s)$ can be introduced as
\begin{equation}
    B(s) \equiv B_0(s) \, = \, \lim_{t\rightarrow 0} B(s,t) \, , 
    \label{e:Bs}
\end{equation}
where the $t\rightarrow 0$ limit, the so called optical point is measured with a finite 
experimental resolution, so in general $B(s)$ is resolution, or $-t$ fit-range, dependent.
The lowest values of $|-t|$ that we analyze in this work correspond to $-t \ge 0.00515$ GeV$^2$ 
at $\sqrt{s} = 7 $ TeV, which is outside of the so-called Coulomb-Nuclear Interference (CNI) region, 
see Fig.~2 of Ref.~\cite{Antchev:2016vpy}. This is why $B(s)$ is called hadronic or nuclear slope, 
as it is defined outside of the range of the CNI effects.

The optical point is also found by extrapolations from the measurements performed in the diffractive 
cone, in the $-t > 0$, experimentally accessible regions. The total cross-section is determined as
\begin{equation}
\sigma_{\rm tot}(s) \equiv 2\,{\rm Im}\, T_{\rm el}(t =0,s) \, .
    \label{e:sigmatot}
\end{equation}
In general, the $(s,t)$ dependent ratio of the real to imaginary parts of the elastic amplitude 
is defined as
\begin{equation}
\rho(s,t)\equiv \frac{{\rm Re}\, T_{\rm el}(s,t)}{{\rm Im}\, T_{\rm el}(s,t)} \, . \label{e:rhost}
\end{equation}
The $t \rightarrow 0$ limit of this ratio is given by 
\begin{equation}
    \rho(s) \equiv \rho_0(s) \, = \, \lim_{t\rightarrow 0} \rho(s,t) \label{e:rhos} \, .
\end{equation}
Technically, $\rho_0$ is measured in the CNI region~\cite{Antchev:2016vpy}, 
with the help of the interference of the well understood Coulomb wave with the nuclear 
(or strong) amplitude. The $t \rightarrow 0$ limit is understood in terms of the finite 
experimental resolution and refers to $\rho$ of the nuclear phase, 
without including the CNI effects. Then the differential cross section 
at the optical point can be written as
\begin{equation}
\frac{d\sigma(s,t)}{dt}\Big|_{t\to 0}=\frac{1+\rho_0^2(s)}{16\pi}\, \sigma_{\rm tot}^2(s) \, .
\label{e:optical-point}
\end{equation}
In the impact-parameter $b$-space, we have 
\begin{eqnarray}\nonumber
	t_{\rm el}(s,b) & = & \int \frac{d^2\mathbf \Delta}{(2\pi)^2}\, 
	e^{-i{\mathbf \Delta}{\mathbf b}}\,T_{\rm el}(s,t) \, , \\
	\Delta & \equiv & |{\mathbf \Delta}|\,, \quad b\equiv|{\mathbf b}|\, ,  \qquad 
	\Delta=\sqrt{|t|} \,. 
	\label{e:Delta}
\end{eqnarray}
Here, the Fourier-transformed elastic amplitude $t_{el}(s,b)$ can be cast 
in the eikonal form as follows
\begin{eqnarray}
	t_{\rm el}(s,b) & = & i\left[ 1 - e^{-\Omega(s,b)} \right] \,,
	\label{e:tel-eikonal}
\end{eqnarray}
where $\Omega(s,b)$ is the so-called opacity function (known also as the eikonal function), 
which is generically complex. The shadow profile function is then defined as
\begin{eqnarray}
	P(s,b) & = & 1-\left|e^{-\Omega(s,b)}\right|^2 \,.
    \label{e:shadow}
\end{eqnarray}
When the real part of the scattering amplitude is neglected, $P(b,s)$ is frequently 
denoted as $G_{\rm inel}(s,b)$, see for example Refs.~\cite{Petrov:2018wlv,Dremin:2013qua,Dremin:2014spa,Dremin:2018urc,Dremin:2019tgm}
for more details.

\section{Odderon}
\label{s:Odderon}
The $pp$ or $p\bar p$ elastic scattering amplitude can be written as a sum or a difference 
of crossing-even and crossing-odd contributions, respectively,
\begin{eqnarray}
T_{\rm el}^{pp}(s,t) & = & T_{\rm el}^{+}(s,t) + T_{\rm el}^{-}(s,t), \\
T_{\rm el}^{p\overline{p}}(s,t) & = & T_{\rm el}^{+}(s,t) - T_{\rm el}^{-}(s,t) , \\
 T_{\rm el}^{+}(s,t) & = & T_{\rm el}^{P}(s,t) + T_{\rm el}^{f}(s,t),\\
 T_{\rm el}^{-}(s,t) & = & T_{\rm el}^{O}(s,t) + T_{\rm el}^{\omega}(s,t) \,.
\end{eqnarray}
The even-under-crossing part consists of the Pomeron and the $f$ Reggeon trajectory, while the odd-under-crossing part contains the Odderon and a contribution from the $\omega$ Reggeon.
A direct way to ``see'' the Odderon in the data is to compare the differential cross-section 
of in elastic $pp$ and $p\bar p$ scattering at TeV energies~\cite{Jenkovszky:2011hu,Ster:2015esa}
since at sufficiently high $\sqrt{s}$ the Reggeon contributions decrease below the experimental errors.
In this case, the Odderon and Pomeron contributions, $T_{\rm el}^{O}(s,t)$ and $T_{\rm el}^{P}(s,t)$, respectively, are found as follows
\begin{eqnarray}
 T_{\rm el}^{P}(s,t) \! & = &  \! \frac{T_{\rm el}^{pp} + T_{\rm el}^{p\overline{p}} }{2} \, \mbox{ \rm if}\, \sqrt{s}\ge 1 \,\, \mbox{\rm TeV} \, , \label{e:Tel-P}   \\
  T_{\rm el}^{O}(s,t) \! & = &  \! \frac{T_{\rm el}^{pp} - T_{\rm el}^{p\overline{p}}}{2}  \, \mbox{ \rm if}\, \sqrt{s}\ge 1 \,\, \mbox{\rm TeV} \,, \label{e:Tel-O} 
\end{eqnarray}
where we have suppressed the $(s,t)$ dependence of the $pp$ and $p\bar p$ scattering amplitudes, 
for the sake of brevity. Indeed, if the $pp$ differential cross sections differ from that 
of $p\bar p$ scattering at the same value of $s$ in a TeV energy domain, then the Odderon contribution 
to the scattering amplitude cannot be equal to zero \cite{Csorgo:2019ewn}, i.e.
     \begin{equation}
        \frac{d\sigma^{pp}}{dt} \neq  \frac{d\sigma^{p\bar p}}{dt} \,\,\, \mbox{ \rm for}\,\, 
        \sqrt{s}\ge 1 \,\, \mbox{\rm TeV}
        \implies 
                T_{\rm el}^O(s,t) \neq 0  \, ,
    \end{equation}
which is considered to be the main Odderon signal. The Odderon contribution can be particularly strong around $t_{\rm dip}$,
as a pronounced diffractive dip (minimum) is seen in elastic $pp$ collisions, while this structure is apparently missing in  $p\bar p$ collisions
in the TeV energy range~\cite{Abazov:2012qb,Antchev:2018rec}.

\subsection{Scaling of the differential cross-section}
\label{ss:scaling}

The low-$|t|$ part of the measured distribution can be 
frequently approximated with an exponential shape, if the experimental errors 
are sufficiently large:
\begin{equation}
    \frac{d\sigma}{dt}(s,t) = A(s) \, \exp\left[ B(s) t\right] \, . \label{e:dsdt-exp}
\end{equation}
The normalization parameter is denoted by $ A(s) = \frac{d\sigma}{dt}(s, t = 0) $ 
and the nuclear slope parameter by $B(s)$ as discussed above. In the diffractive cone, 
$A(s) = B(s) \sigma_{\rm el}(s) $, and the nearly exponential differential cross-sections
can be scaled to a universal scaling function defined as \cite{Csorgo:2019ewn}
\begin{eqnarray}
    H(x) & \equiv  & \frac{1}{B(s) \sigma_{\rm el}(s)} \frac{d\sigma}{d t} \,, \label{e:H(x)}\\
    x    & = &  - t B(s) \, . \label{e:x}
\end{eqnarray}
At low-$|t|$ this scaling function is approximately 
written as $H(x) = \exp(-x)$. Thus, $H(x)$ scales out the trivial $s$ dependencies 
of $B(s)$ and $\sigma_{\rm el}(s)$ from the differential cross-sections 
in the diffractive cone. In the exponential approximation,
\begin{eqnarray}
    A(s) & = & B(s) \, \sigma_{\rm el}(s) \, = \, \frac{1+\rho_0^2(s)}{16 \, \pi}\, 
    \sigma_{\rm tot}^2(s) \, , \label{e:Asigma}\\
    B(s) & = & \frac{1+\rho_0^2(s)}{16 \, \pi }\, 
    \frac{\sigma_{\rm tot}^2(s)}{\sigma_{\rm el}(s)} \, , \label{e:Bsigma}
\end{eqnarray}
thus $H(x)$ scales out the $s$-dependence, 
if it arises only due to the $s$-dependence of the total cross-section 
$\sigma_{\rm tot}(s)$ and the real-to-imaginary ratio $\rho(s)$. 

In Ref.~\cite{Csorgo:2019ewn} we have shown that our $H(x)$ scaling, defined as above, 
is valid not only in the diffractive cone but, surprisingly, also at the diffractive 
minimum and maximum of elastic proton-proton collisions: the $H(x,s)$ is $s$-independent 
in the ISR energy range of $\sqrt{s} = 23.5 $ - $62.5$ GeV. In addition, the $H(x)$ scaling 
was shown to be also valid, within statistical errors, in the LHC energy range of 
$\sqrt{s} = $ $2.76 - 7.0$ TeV, but with a scaling function
that is significantly different from the one at ISR energies~\cite{Csorgo:2019ewn}.

From the $s$-independence of the $H(x,s)$ at the lower LHC energies as well as from Eqs.~(\ref{e:H(x)},\ref{e:x}) a new Odderon signature follows~\cite{Csorgo:2019ewn}:
If the $H(x,s)$ scaling function for $pp$ collisions differs from that of $p\bar p$ in 
a given $s$ domain, where $H(x,s) \equiv H(x)$ is $s$-independent for $pp$ collisions, 
then the differential cross-sections in that interval of $s$ cannot be equal 
for $pp$ and $p\bar p$ collisions, either. Hence, the Odderon amplitude cannot be 
equal to zero in that $s$-range:
     \begin{eqnarray}
     \nonumber
         \mbox{\rm If}  \, H(x,s_1) \, \mbox{\rm for } \, pp & \neq &   
         H(x,s_2) \, \mbox{ \rm for} \, p\bar p \, \\
        \nonumber
     \null &\mbox{\rm for} &   1 \, \lessapprox \sqrt{s_1}, \sqrt{s_2} 
     \lessapprox 8 \, \mbox{\rm TeV} \\ 
        \null &\implies &
                T_{\rm el}^O(s,t) \neq 0  \, .
    \end{eqnarray}

\begin{figure*}[hbt]
\begin{center}
\begin{minipage}{1.0\textwidth}
 \centerline{\includegraphics[width=0.48\textwidth]{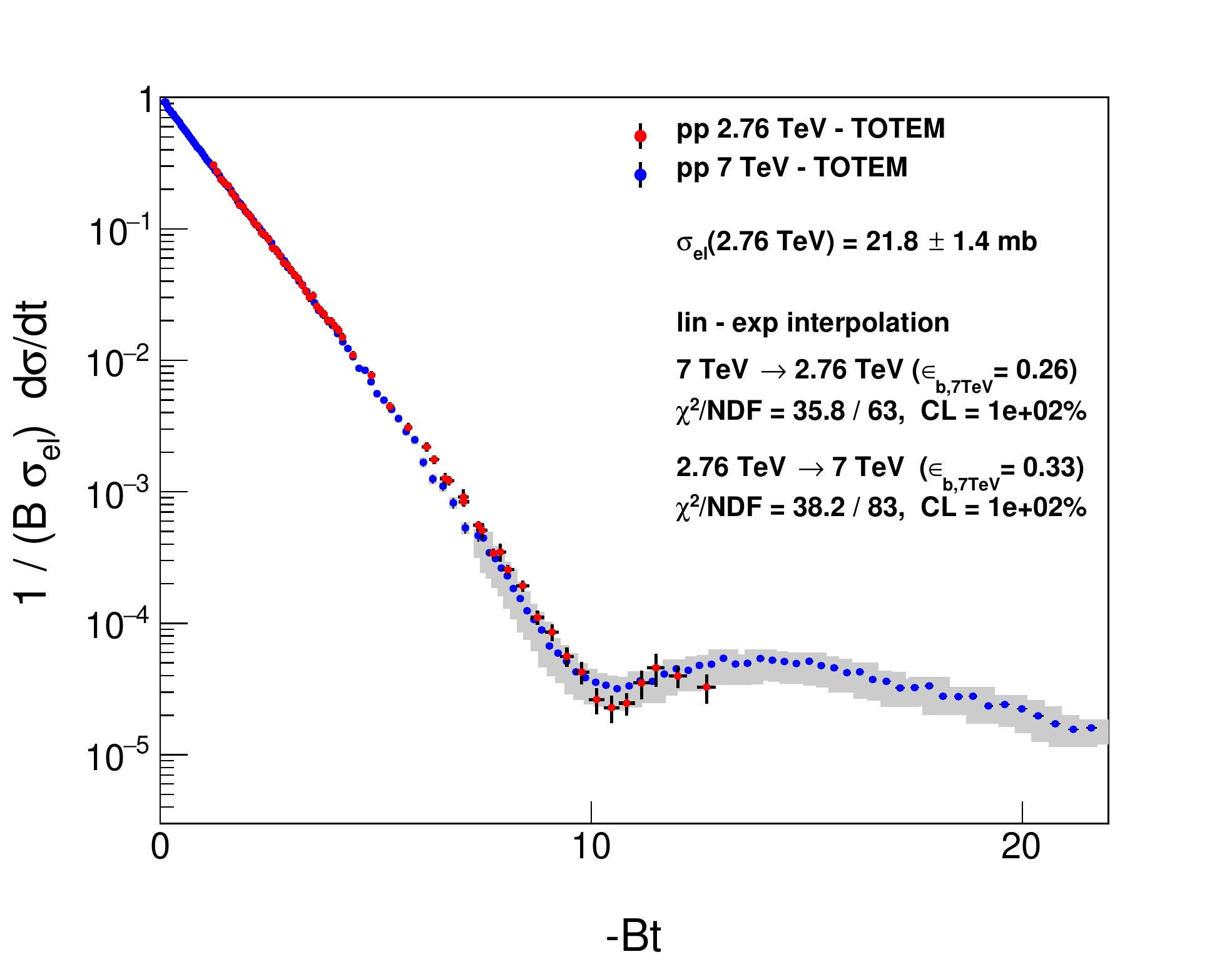}
 \includegraphics[width=0.48\textwidth]{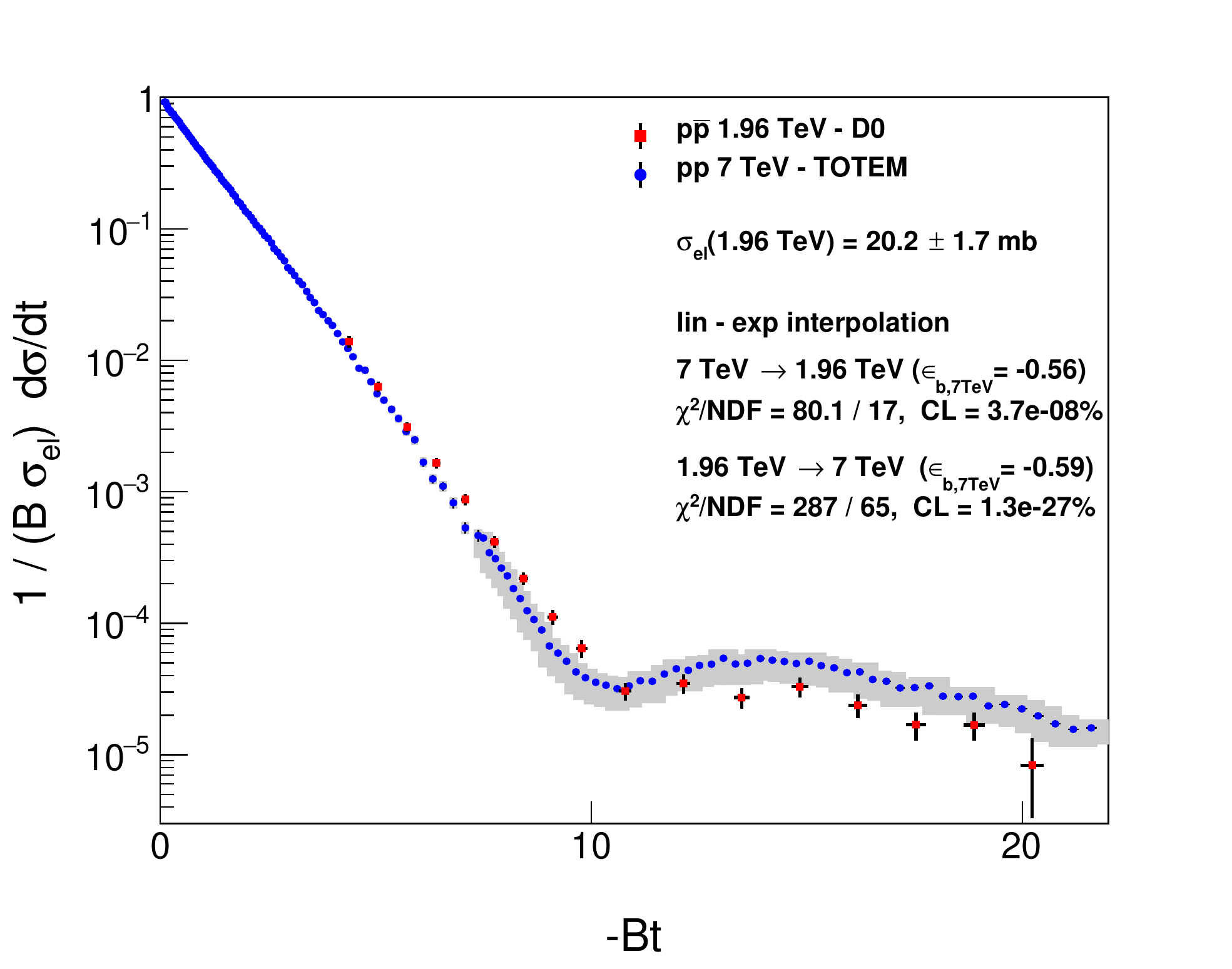}}
\end{minipage}    
\vspace{-0.6truecm}
\end{center}
\caption{ 
Left panel: For proton-proton elastic scattering, the $H(x,s) = \frac{1}{B \sigma_{\rm
el}}\frac{d\sigma}{dt} $ scaling function is energy independent in terms of $x = - t B$ 
at the TeV energy scale, or specifically, in the $\sqrt{s} = 2.76$ - $7$ TeV energy range. TOTEM preliminary data
at $\sqrt{s} = 8 $ TeV~\cite{Kaspar:2018ISMD} also follow this scaling law, while scaling violations are observed
at $\sqrt{s} = 13$ TeV~\cite{Csorgo:2019ewn}.
Right panel: A statistically significant difference is found between the $H(x,s)$ scaling 
functions for elastic $pp$ collisions at $\sqrt{s} = 7.0$ TeV and that of $p\bar p$ collisions at
$\sqrt{s} = 1.96$ TeV. On both panels, vertical lines stand for the type A (point-to-point fluctuating)
errors, while the horizontal lines indicate the point-to-point fluctuating uncertainty of the horizontal
errors. Grey vertical bars stand for the type B (point-to-point changing, but overall correlated) errors
and the width of this grey bars correspond to the type B errors in the horizontal $(x)$ direction. Larger and smaller boxes correspond to two different measurements at $\sqrt{s} $ $=$ $7$
TeV, detailed in Refs.~\cite{Antchev:2011zz,Antchev:2013gaa}, respectively.
}
\label{fig:H(x)-Odderon}
\end{figure*}

The left panel of Fig.~\ref{fig:H(x)-Odderon} indicates that for elastic proton-proton collisions,
the $H(x,s)$ scaling function of the differential cross-section is independent, within statistical errors, of the colliding energy for $\sqrt{s} = 2.76$ and $7.0$ TeV, at the confidence level 
of $CL = 99$ \%. Similar results are obtained for the TOTEM preliminary large $-t$ differential
cross-sections at $\sqrt{s} $ $=$ $8$ TeV~\cite{Kaspar:2018ISMD}. Given that in addition to the
statistical errors, there are significant systematic errors present as well, this result 
indicates that the validity of the $s$-independence of the $H(x) \equiv H(x,s)$ scaling 
for $pp$ collisions is greater than the $2.76 \le \sqrt{s} \le 8.0$ TeV region.
Within {\it systematic} errors, indicated by vertical bars with widths proportional 
to the horizontal systematic error on $x=-tB$, the confidence level of the agreement 
of the $H(x,s)$ scaling function for the two datasets at $\sqrt{s} = 2.76$ and $7.0 $ TeV 
is 100 \%. In Fig. ~\ref{fig:H(x)-Odderon}, the regions with smaller and larger boxes 
correspond to two different datasets, measured in two different acceptance regions 
\cite{Antchev:2011zz,Antchev:2013gaa}.

Quantitatively, we have found a statistically significant Odderon signal in the comparison 
of the $H(x,s)$ scaling functions of elastic $pp$ collisions at $\sqrt{s} = 7.0 $ to that 
of $p\bar p$ collisions at $\sqrt{s} = 1.96$ TeV. On the right panel of Fig.~\ref{fig:H(x)-Odderon}, 
we compare the $H(x,s)$ scaling functions of elastic $pp$ collisions at $\sqrt{s}$ $=$ $7.0$ TeV
with that of the elastic $p\bar p$ collisions at $\sqrt{s} = 1.96 $ TeV. These scaling functions are
statistically significantly different. The confidence level of their agreement is maximum $CL = 3.7
\times 10^{-8}$ \%, corresponding to a statistically significant,  of at least $6.26$ $\sigma$  Odderon signal. 
This difference is larger, than the 5$\sigma$ threshold, required for a discovery in particle physics.
The advantages of our method, with respect to comparing the cross sections directly include the scaling out of
the $s$-dependencies of $\sigma_{\rm tot}(s)$, $\sigma_{\rm el}(s)$, $B(s)$ and $\rho(s)$, as well as
the normalization of the $H(x)$ scaling function that cancels the point-to-point correlated and $t$-independent normalization errors.

\section{Proton holography}
\label{s:holography}
A new possibility for phase reconstruction in elastic proton-proton scattering 
is based on our detailed analysis of elastic $pp$ collisions data in the dip region 
at ISR and LHC energies~\cite{Csorgo:2018uyp,Csorgo:2018ruk,Csorgo:2019zcj,Csorgo:2019egs,Csorgo:2019ewn}. To support this conjecture, we present some qualitative considerations 
as well as the results of a model-independent study, and also discuss the currently known 
limitations of this promising new method.

\begin{figure*}[hbt]
\begin{center}
\begin{minipage}{1.0\textwidth}
 \centerline{\includegraphics[width=0.48\textwidth]{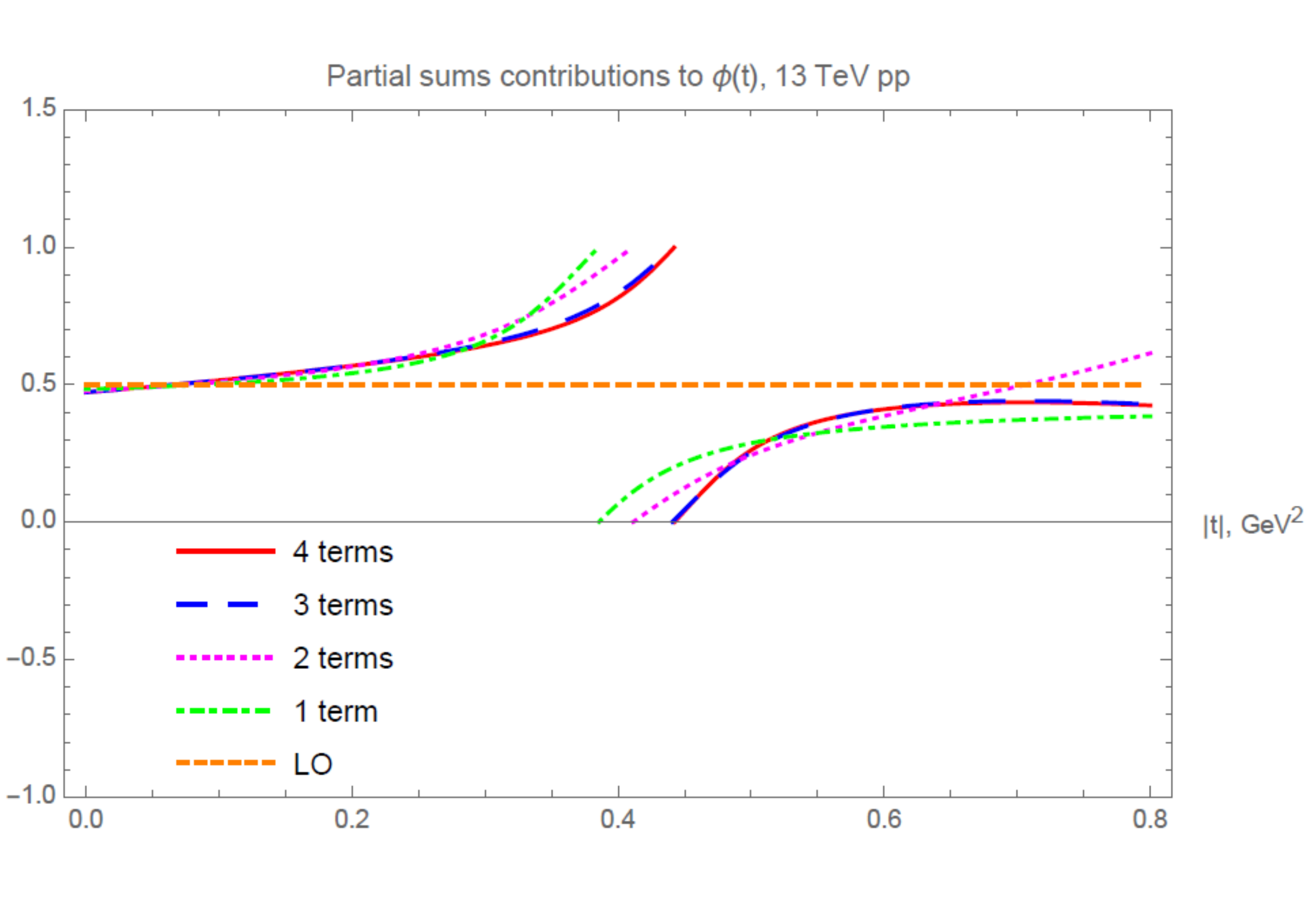}
 \includegraphics[width=0.48\textwidth]{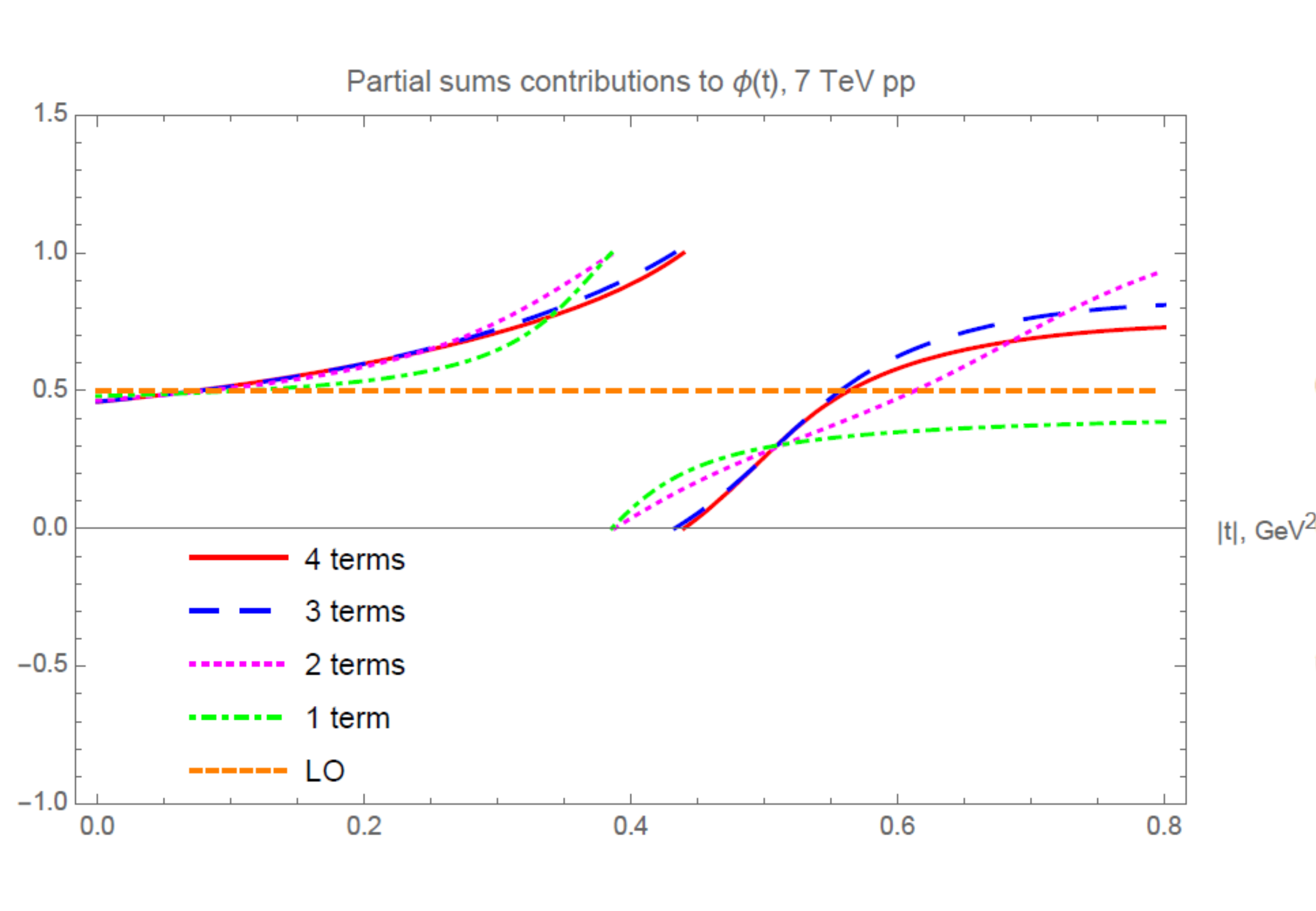}}
\end{minipage}  
\vspace{-0.6truecm}
\end{center}
\caption{ 
Left: For $pp$ elastic scattering at $\sqrt{s}$ $=$ $13.0$ TeV, 
the nuclear phase reconstruction based on a 4th order Levy expansion converges 
up to $-t $ $\lessapprox $ $ 0.9$ GeV$^{2}$. Right: The 
phase reconstruction at $\sqrt{s}$ $ = $ $ 7.0$ TeV converges up to $-t \lessapprox 0.5$ GeV$^{2}$, as detailed in~\cite{Csorgo:2019zcj}.
}
\label{fig:p-holography}
\end{figure*}

As obvious, the phase reconstruction from the measurement of the differential cross-section 
of elastic scattering, in general and without additional
information, is not possible. For example, if the measurements
are limited to the diffractive cone, outside of the CNI region, but 
stopping well before the diffractive minimum-maximum (dip and bump) structure, 
the phase reconstruction from a structureless, nearly exponential cone is clearly meaningless 
and impossible. These limiting cases were discussed for example in Appendices C and D of
Ref.~\cite{Csorgo:2018uyp}. However, the observation of a diffractive minimum, followed 
by a diffractive maximum corresponds to the observation of an interference pattern, 
one of the conditions of phase reconstruction and holography. Such an interference pattern 
is clearly visible as a function of the deflection angle in Fig.~1 of Ref.~\cite{Csorgo:2019fbf}, originally from Ref.~\cite{Antchev:2018edk}.

Recently, we have discussed model-independent, whole $-t$ range fits to the differential
cross-section in Appendix A and B of Ref.~\cite{Csorgo:2018uyp} for $pp$ and for $p\bar p$ 
elastic collisions, respectively. A closer inspection of these fits indicate that at the ISR 
energy range, the total cross-section is reproduced reasonably well, while the $\rho_0$ measurements 
are reproduced only in one case, at $\sqrt{s} = 44.7$ GeV, within two standard deviations.
Given that we did {\it not} utilize any data in the CNI region, the bad reproduction of several 
of the measured $\rho_0$ values were not surprising. As discussed in Ref.~\cite{Csorgo:2018uyp}, 
in the case of the very well measured TOTEM data at $\sqrt{s} = 13$ TeV, we were able to reproduce 
within experimental errors also the measured value of $\rho_0$, and we have obtained similarly good
quality preliminary results at $\sqrt{s} = 7 $ TeV as well. 

We started to wonder, whether or not such a good quality $\rho_0$ determination from outside 
the CNI region is a coincidence only: it is possible that there is a deeper reason for it.
As a first step, we have shown in Ref.~\cite{Csorgo:2018uyp}, that {\it within} the technique 
of the model-independent L\'evy series expansion, the reconstruction of the $t$-dependent 
phase or equivalently the determination of the $-t$ dependent $\rho(t)$ function is unique,
similarly to how the coefficients of a Taylor-series can be uniquely determined if the measurement 
of the Taylor-expanded function is precise enough. Of course, behind this reconstruction lies a powerful
physical assumption about the L\'evy processes driving the elastic scattering in QCD which
dictates a very specific form of the elastic amplitude, in the form of generalised L\'evy series.
However, other methods may result in other $\phi(t)$ functions. We are currently working on these
subtleties of the $-t$ dependent phase reconstruction by comparing the results of
Ref.~\cite{Csorgo:2018uyp} with other, model-dependent efforts.

In this work, we thus start to explore some of the basic details of $t$-dependent phase 
reconstruction. We ask the following question: what are those special experimentally testable, 
but mathematically well defined conditions that are sufficient for an experimentally validated 
phase reconstruction in elastic $pp$ or $p\bar p$ scattering?
Let us recall two examples here about how such a phase reconstruction might work in elastic $pp$ scattering.

The first example deals with phase reconstruction at the ISR energies. Such a reconstruction of the
nuclear phase from the ISR data was performed in Ref.~\cite{Nagy:1978iw} utilizing the Phillips-Barger
model~\cite{Phillips:1974vt}. This model was shown to be well suited to describe the ISR data in the
$\sqrt{s} = 23.5$ - $62.1$ GeV region with statistically acceptable, good quality fits in the 
$1.05 \le -t \le 2.5$ GeV$^2$ interval. Table 3 of Ref.~\cite{Nagy:1978iw} reported the values of
$|\epsilon|$, the modulus square of the phase difference from $\pi$, with values small but close 
to zero in the range of (0.06 $\pm$ 0.06) -- (0.34 $\pm$ 0.10). These values suggested that the imaginary
part of the scattering amplitude approximately vanishes near the diffractive minimum, and the real part
is dominant at the dip region, but due to the $\cos(\phi)  = \cos(\pi \pm \epsilon) \approx 1 + 0.5
\epsilon^2$ type of Taylor expansion, the sign of the phase difference, $\epsilon$ cannot be determined
from the experimental data. Assuming that a crossing-odd contribution is negligible at the ISR energies,
a more detailed, $t$-dependent reconstruction of $\rho(t) = {\mathcal Re} T/{\mathcal Im} T$ is reported
in Fig. 22 of Ref.~\cite{Nagy:1978iw}.

The Reggeized Phillips-Barger model~\cite{Phillips:1974vt} gave good quality fits to the ISR and 
LHC data in larger $-t$ intervals as well, starting from $-t$ $\ge$ 0.3 GeV$^{2}$ and in 2011 
it was used to study the Odderon effect in the elastic $pp$ scattering data at LHC
energies~\cite{Jenkovszky:2011hu}. However, this model at small values of $-t$ $\leq $ 0.3 GeV$^{2}$ 
did not give a statistically acceptable data description, hence this model cannot be used to connect 
the phase measurement at the dip region of $-t \approx 0.5$ GeV$^2$ with CNI measurements
at the optical point $t\approx 0$ GeV$^2$, so this model cannot be used to resolve 
the Cul-de-Sac of elastic scattering~\cite{Dremin:2019tgm}.
However, the model-independent L\'evy series~\cite{Csorgo:2018uyp} is able to do this job. 
As far as we know, this is the only method so far, that is able to describe the elastic $pp$ 
collisions at the currently highest energy of $\sqrt{s}$ $=$ 13 TeV~\cite{Antchev:2018edk} 
with a statistically acceptable, 2\% confidence level. Within this model-independent approach 
the $t$-dependent phase reconstruction appears to be unique~\cite{Csorgo:2018uyp}. Increasing 
the order of the series, the domain of the convergence starts from $t\approx 0$ and 
is increasing gradually. The convergence region for the phase of the fourth order L\'evy series 
is  $0\le -t \lessapprox 0.5$ and $0.9 $ GeV$^2$ at $\sqrt{s} = 7 $ and $13$ TeV, respectively, as shown on Fig.~\ref{fig:p-holography}.

The left panel of Fig.~\ref{fig:p-holography} indicates that for proton-proton elastic scattering 
at $\sqrt{s}$ $=$ $13.0$ TeV, the nuclear phase reconstruction based on a 4th-order Levy expansion
converges up to $-t $ $\lessapprox $ $ 0.9$ GeV$^{2}$. The right panel of the same figure indicates 
similar results but at $\sqrt{s}$ $=$ $7.0$ TeV. These data are less precise as compared 
to the measurements at $\sqrt{s} = 13.0 $ TeV, correspondingly the domain of convergence 
of the phase reconstruction is smaller. Generically, the greater the experimental precision, 
the larger the domain of convergence of this method. These plots suggest that the reconstruction 
of the $t$-dependent phase is possible with the help of this L\'evy series technique, 
providing a new kind of model-independent proton holography. This is our second example, 
detailed below.

One of the most intriguing properties of elastic $pp$ collisions
at the ISR energy range of $\sqrt{s} = 23.5 - 62.1$ GeV and at the LHC energy range
of $\sqrt{s} = 2.76  - 13. 0 $ TeV is the existence of a unique, single diffractive minimum,
and the existence of a nearly exponential diffractive cone not only at very low $|t|$
but also a similar, secondary diffractive cone that follows after the diffractive
maximum~\cite{Csorgo:2019egs}. From measurements of the CNI region, reviewed recently 
for example in Fig.~15 of Ref.~\cite{Antchev:2017yns}, we also know that in each of 
these measurements, $\rho_0 \le 0.15$ that implies that close to $t\approx 0$, the real 
part of the scattering amplitude is, as compared to the imaginary part, relatively small.

By means of the optical theorem, a total cross section measurement uniquely determines 
the imaginary part of the elastic amplitude at $t=0$, as indicated by Eq.~(\ref{e:sigmatot}). 
This implies that not only the imaginary but also the $\rho_0^2$ or the squared value 
of real part of the forward scattering amplitude can also be uniquely determined at 
the optical point, $t=0$. Indeed, using the diffractive cone approximation and
Eqs.~(\ref{e:Asigma}) and (\ref{e:Bsigma}), from the values of $A(s)$, $B(s)$ and 
$\sigma_{\rm tot}(s)$, the values of $\sigma_{\rm el}(s)$ and $\rho^2_0(s)$ are determined 
uniquely. From such a determination of $\rho_0^2(s)$, the value of the phase of 
the scattering amplitude at the optical point, $\phi(t=0)$ is fixed to two possible 
values by measurements. However, to select one of these values unambiguously, further 
information about the CNI interference has to be utilized. This dual degeneracy is related 
to the fact that the sign of $\rho_0$ cannot be determined from measurements of $\rho_0^2$, 
such that both possible values, corresponding to  $\pm \sqrt{\rho_0^2(s)}$, are allowed. 
These two different allowed phases correspond to the object and the conjugated object 
wave reconstruction in holography.

One can determine the nuclear phase at the dip $t_{\rm dip}$, too, where the differential 
cross-section of elastic proton-proton collisions has a diffractive minimum. At this point, 
the imaginary part of the elastic scattering amplitude approximately vanishes,
and the differential cross-section is dominated in the dip region by the real part of the forward
scattering amplitude. Thus, the nuclear phase is approximately an integer multiple of $\pi$ close 
to the dip region. This is a degeneracy similar to the $\rho_0$ sign problem, mentioned 
in the previous paragraph, and to the object beam and the conjugated object beam reconstruction 
in holography, mentioned in the Introduction. This problem has been called the Cul-de-Sac 
of elastic scattering in Ref.~\cite{Dremin:2019tgm}: at the diffractive minimum, 
only the modulus but not the sign of the real part can be uniquely determined, 
hence the nuclear phase is either $0$ or $\pi$, up to small corrections.

The fact that at $\sqrt{s} = 13$ TeV, the phase at the dip connects analytically to the phase 
measured in the CNI region indicates, that with the help of the L\'evy expansion method, the
nuclear phase of the elastic scattering amplitude can be realistically determined, at least 
in certain specific cases. A posteriori, our method is validated by the excellent reproduction 
of the $\rho_0$ value measured in the CNI region in elastic $pp$ collisions at $\sqrt{s} = 13$ TeV 
by the TOTEM Collaboration~\cite{Antchev:2017yns}. Fits detailed in Appendices A and B of
Ref.~\cite{Csorgo:2018uyp} indicate that indeed the 4th and 3rd order L\'evy expansions 
reproduce well measured values of $\rho_0$ not only at $\sqrt{s} = 13 $ TeV but at lower LHC 
energies as well, if the fit quality is satisfactory. However, one should put a pull on $\rho_0$ 
in the definition of $\chi^2$ as the sign problem at the dip cannot be resolved otherwise. 
The two different signs of the phase near the dip can lead to two different analytic
continuation of the phase to $t=0$ but only one of these continuations lead to the correct 
value of $\rho_0$.

\section{Summary}
\label{s:summary}
We have described a new Odderon signal~\cite{Csorgo:2019ewn}, the validity of the $H(x)$ scaling 
for $pp$ collisions and its violation in $p \bar p$ collisions in the few TeV energy range
shown and quantified in the right panel of Fig.~\ref{fig:H(x)-Odderon}.
The statistical significance of this Odderon signal is found to be at least $6.26\sigma$.

We have also discussed the related topic of proton holography or the $t$-dependent reconstruction
of the nuclear phase. If we assume that this nuclear phase $\phi(t)$ is an analytic and continuous
function of $t$, and the differential cross-section of elastic scattering data is determined
with great precision, the two different approximate values of the phases at the dip, $0$ and $\pi$ 
can be analytically and continuously extrapolated to $t\rightarrow 0$. Furthermore, our analysis 
with L\'evy series performed in Ref.~\cite{Csorgo:2018uyp} suggested that if these two extrapolated
values are within the errors of the extrapolations significantly different from one another, and 
only one of them is consistent with the determination of the $\rho_0$ using the CNI methods, then 
the $t$-dependence of the nuclear phase can be reconstructed not only at a given value of $t=0$ 
or $t=t_{\rm dip}$ but in the entire $t$ region, where such an extrapolation is possible 
and convergent.

The search for a sufficient and necessary condition that would make the $t$-dependent 
phase reconstruction unique not only within, but also outside the L\'evy series method 
is still ongoing at the time of closing this manuscript.

\section*{Acknowledgments}
We thank I. Vitev and co-organizers for an inspiring and successful meeting. We acknowledge stimulating
discussions with G. Gustafson, Y. Hatta, L. L\"onnblad, and M. \v{S}um\-bera as well as with members 
of the TOTEM Collaboration. T. Cs. and A. S. were partially 
supported by the NKFIH grants No. FK-123842, FK-123959 and K-133046, and by 
the EFOP 3.6.1-16-2016-00001 grant (Hungary). 
R. P. is partially supported by Swedish Research Council grants No. 621-2013-4287 
and 2016-05996, by ERC H2020 grant No 668679, as well as 
by the Ministry of Education, Youth and Sports of the Czech Republic project LTT17018. 
Our collaboration was supported by THOR, the EU COST Action CA15213. 
%
\bibliography{p-holography}
\thispagestyle{empty} 
\clearpage
\vfill\eject

\end{document}